\documentclass[12pt,epsf]{article}
\usepackage{graphicx,amsmath,amssymb,booktabs,enumerate,url,color}
\usepackage{subfigure}
\begin{document}

\def\a{\alpha}
\def\b{\beta}
\def\c{\varepsilon}
\def\d{\delta}
\def\e{\epsilon}
\def\f{\phi}
\def\g{\gamma}
\def\h{\theta}
\def\k{\kappa}
\def\l{\lambda}
\def\m{\mu}
\def\n{\nu}
\def\p{\psi}
\def\q{\partial}
\def\r{\rho}
\def\s{\sigma}
\def\t{\tau}
\def\u{\upsilon}
\def\v{\varphi}
\def\w{\omega}
\def\x{\xi}
\def\y{\eta}
\def\z{\zeta}
\def\D{\Delta}
\def\G{\Gamma}
\def\H{\Theta}
\def\L{\Lambda}
\def\F{\Phi}
\def\P{\Psi}
\def\S{\Sigma}

\def\o{\over}
\def\beq{\begin{eqnarray}}
\def\eeq{\end{eqnarray}}
\newcommand{\gsim}{ \mathop{}_{\textstyle \sim}^{\textstyle >} }
\newcommand{\lsim}{ \mathop{}_{\textstyle \sim}^{\textstyle <} }
\newcommand{\vev}[1]{ \left\langle {#1} \right\rangle }
\newcommand{\bra}[1]{ \langle {#1} | }
\newcommand{\ket}[1]{ | {#1} \rangle }
\newcommand{\EV}{ {\rm eV} }
\newcommand{\KEV}{ {\rm keV} }
\newcommand{\MEV}{ {\rm MeV} }
\newcommand{\GEV}{ {\rm GeV} }
\newcommand{\TEV}{ {\rm TeV} }
\def\diag{\mathop{\rm diag}\nolimits}
\def\Spin{\mathop{\rm Spin}}
\def\SO{\mathop{\rm SO}}
\def\O{\mathop{\rm O}}
\def\SU{\mathop{\rm SU}}
\def\U{\mathop{\rm U}}
\def\Sp{\mathop{\rm Sp}}
\def\SL{\mathop{\rm SL}}
\def\tr{\mathop{\rm tr}}

\def\IJMP{Int.~J.~Mod.~Phys. }
\def\MPL{Mod.~Phys.~Lett. }
\def\NP{Nucl.~Phys. }
\def\PL{Phys.~Lett. }
\def\PR{Phys.~Rev. }
\def\PRL{Phys.~Rev.~Lett. }
\def\PTP{Prog.~Theor.~Phys. }
\def\ZP{Z.~Phys. }


\baselineskip 0.7cm

\begin{titlepage}
YITP-12-5

\vskip 1.35cm
\begin{center}
{\large \bf Gravitational Wave Probe of High Supersymmetry Breaking Scale
}
\vskip 1.2cm
Ryo Saito${}^1$ and Satoshi Shirai${}^{2,3}$
\vskip 0.4cm

{\it

$^1$ Yukawa Institute for Theoretical Physics, Kyoto University,\\
Kyoto 606-8502, Japan\\
$^2$Department of Physics, University of California, \\Berkeley, CA 94720\\
$^3$ Theoretical Physics Group, Lawrence Berkeley National Laboratory, \\Berkeley, CA 94720

}

\vskip 1.5cm

\abstract{
A supersymmetric standard model with heavier scalar particles is very interesting from 
various viewpoints, especially Higgs properties.
If the scalar mass scale is ${\cal O}(10-10^4)$ TeV, the standard model-like Higgs with mass around 125 GeV,
which is implied by the recent LHC experiments, is predicted. 
However this scenario is difficult to be directly tested with collider experiments.
In this paper, we propose a test of this scenario by using observations of primordial gravitational waves (GWs). The future GW experiments such as DECIGO and BBO can probe the scalar mass around ${\cal O}(10^3-10^4)$ TeV, which is preferred from the Higgs mass about 125 GeV, if the primordial GWs have large amplitude.
}
\end{center}
\end{titlepage}

\setcounter{page}{2}

\section{Introduction}
A supersymmetric (SUSY) standard model (SSM) is a very appealing model beyond the standard model (SM).
An important prediction of the SSM is lightness of SM-like Higgs particle.
The recent observation by ATLAS and CMS collaborations suggests the existence of the Higgs particle with mass around 125 GeV \cite{ATLAS,CMS}.
To realize the Higgs mass about 125 GeV, the SSM can be strongly constrained: heavy scalar top quark and/or very large $A$-term of the scalar tops are required
in the case of the minimal SSM (MSSM) \cite{Higgs, Heinemeyer:2011aa,Arbey:2011ab}.

Among them, heavy scalar scenarios are interesting from various viewpoints.
 With the small $A$-term of the stops, stop mass around ${\cal O}(10-10^4)$ TeV can successfully explain the Higgs mass about 125 GeV \cite{Arbey:2011ab, Giudice:2011cg, Ibe:2011aa}.
In addition to the Higgs mass, this scenario is successful in the SUSY flavor and CP problem and/or the cosmological gravitino problem \cite{Kawasaki:2008qe} as discussed in the context of the split SUSY \cite{ArkaniHamed:2004fb,Giudice:2004tc,ArkaniHamed:2004yi} or the anomaly mediated SUSY breaking (AMSB) models \cite{Randall:1998uk, Giudice:1998xp}.
Motivated with such a scenario with heavy scalar mass, phenomenological consequences are discussed in Refs. \cite{Ibe:2011aa,Moroi:2011ab}.

However, this scenario is very difficult to be directly tested with collider experiments.
This is because the scalar particles are too heavy to be produced at colliders, even if
the gauginos are light enough to be produced at colliders.
However once the gluino-like particle is discovered, 
there is a chance to probe the scalar sector.
The gluino decay is very sensitive to the scalar sectors.
For example, if the scalar mass scale is larger than about $10^{3-4}$ TeV,
the gluino has a sizable lifetime.
Therefore, an R-hadron or a displaced vertex of a long-lived gluino can be a probe of such a heavy scalar mass
 \cite{ArkaniHamed:2004fb,Toharia:2005gm}.
Although this is an interesting possibility, it is still indirect evidence for the heavy scalar SUSY scenario.
Thus, it is very important to investigate the independent way to probe the scalar sector.

In this paper, we discuss testability of this scenario with the next generation projects of gravitational wave (GW) experiments as Deci-hertz Interferometer Gravitational Wave Observatory (DECIGO) \cite{DECIGOWeb, Seto:2001qf, Kawamura:2011zz} and Big Bang Observer (BBO) \cite{BBO}. 
A main target of the next generation GW projects is the primordial GWs generated in the inflationary era \cite{Starobinsky:1979ty}.
The observations of the primordial GWs can provide information not only on inflation but also on the thermal history of the early Universe \cite{TH, Watanabe:2006qe}. 
 In the case of the heavy scalar top mass scenario,  it is natural to expect that the other SSM scalar particles have masses of the same order of magnitude. 
 Therefore, the sudden change of relativistic degrees of freedom as much as $\sim 100$ is expected at temperature around the scalar mass, which leads to a step at the corresponding frequency in the amplitude of the GWs \cite{Watanabe:2006qe, Chiba:2007kz}. 
 The relevant energy range ${\cal O}(10^2-10^4)$ TeV corresponds to the frequency band of DECIGO/BBO, ${\cal O}(0.001-0.1)$ Hz. 
 We then investigate whether DECIGO/BBO can detect a step in their band caused by the change in the relativistic degrees of freedom using the correlation analysis techniques \cite{CA}. 
 By measuring frequency dependence of the primordial GWs in detail, we can obtain information of very high-mass-scale particle beyond the reach of the collider experiments, which gives a good test of the SUSY models.

\section{Heavy scalar scenario and cosmology}
 We describe here the Higgs properties and cosmological consequences of our setup with high SUSY breaking scale.
In the low-energy below the heavier SSM scalar masses, there is a SM-like Higgs scalar doublet $H$. 
The relevant Higgs potential is similar to the SM one and written as
	\begin{align}\label{eq:higgs_potential}
		V(H) = \frac{\lambda}{2}(HH^{\dagger} - v^2)^2,
	\end{align}
where $v = 174$ GeV is the Higgs VEV and the Higgs mass is provided as $m_h = \sqrt{2 \lambda} v$.
At tree-level in the MSSM, the $\lambda$ comes from $D$-term potential and is related to the gauge couplings as,
	\begin{align}\label{eq:4points}
	\lambda = \frac{1}{4}(g_2^2 + \frac{3}{5}g_1^2)\cos^2(2\beta).
	\end{align}
 Hence, the value of the Higgs mass cannot be arbitrarily large. Instead, the Higgs mass is bounded above as,
	\begin{align}\label{eq:higgs_tree}
	m_h = m_Z \cos(2\beta) \lsim 91~{\rm GeV},
	\end{align}
at tree-level where $\beta$ is defined by $\tan\beta = v_u/v_d$. 
This value is quite small compared to plausible Higgs mass $\sim 125$ GeV. A large value of the Higgs mass is realized by including contributions from the radiative corrections, $\Delta m_h^2$, which are controlled with the SUSY breaking scale \cite{Higgs}.
Although the detailed value depends on SUSY breaking parameters and $\tan \beta$,
in the case of small $A$-term of the stops, which is plausible in the context of the split SUSY and AMSB scenarios, 
${\cal O}(10-10^4)$ TeV stop mass provides appropriate Higgs mass.
In such a case, it is natural to expect that the masses of MSSM scalar other than stop are also heavy, $m_{\rm scalar} \simeq m_{\rm stop}$.
For example, in the case of the AMSB models, the scalar masses are controlled with the gravitino mass $m_{3/2}$
and, hence, all the scalar masses are likely as large as the stop mass. 
Therefore, there are expected to be nearly a hundred of particles other than the SM above the scalar mass scale, $m_{\rm scalar}$.
Hereafter, we assume that all the MSSM scalar particles have universal scalar mass $m_0$ and
that the $\mu$-term is also large: $\mu = m_0$. 
These assumptions are just for simplicity. Even if we consider the variation in the scalar masses, the results presented below do not change so much.

\begin{figure}
\begin{center}
{\includegraphics[clip]{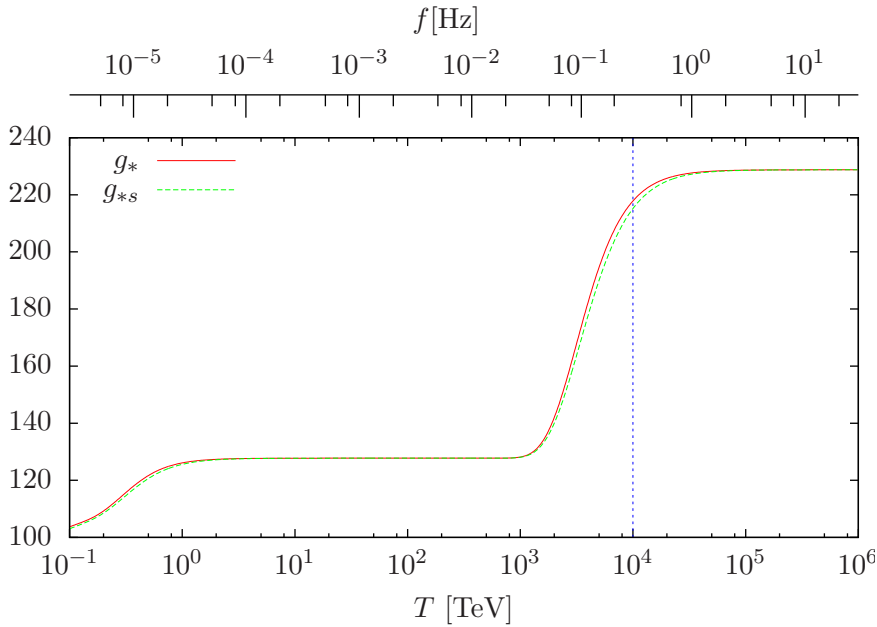}}
\caption{Evolution of $g_*$ and $g_{*s}$ as functions of temperature $T$. We have set $m_{\rm gaugino} = 1$ TeV and $m_0 = 10^4$ TeV. 
}
\label{fig:g}
\end{center}
\end{figure}

Then let us consider the cosmological history in this scenario. In this paper, we focus on a fact that the effective number of relativistic degrees of freedom changes at the scalar mass scale, $T \sim m_0$. 
We define the effective number of relativistic degrees of freedom $g_*$ and $g_{*s}$ as,
	\begin{align}\label{eq:gstar}
	\rho = \frac{\pi^2 g_*(T)}{30} T^4,\\
	s = \frac{2\pi^2 g_{*s}(T)}{45} T^3,
	\end{align}
where $\rho$ and $s$ are the energy density and the entropy density, respectively.
The evolution of $g_*$ and $g_{*s}$ are determined by the MSSM mass spectrum.
In Fig. \ref{fig:g}, we have shown an example of $g_*$ and $g_{*s}$
when $m_{\rm gaugino}=1$ TeV and $m_0 = 10^4$ TeV. 
From the figure, we can see that the relativistic degrees of freedom increase by $100$ at temperature corresponding to the scalar mass scale, $T \simeq 10^4~{\rm TeV}$.

The $g_{\ast}$ change discussed here causes features in the GW spectrum \cite{Watanabe:2006qe, Chiba:2007kz}. In the next section, we investigate the detectability of this $g_{\ast}$ change by the next generation GW experiments as DECIGO \cite{DECIGOWeb, Kawamura:2011zz} and BBO \cite{BBO}.

\section{Detecting $g_{\ast}$ change in the gravitational wave background spectrum}

\subsection{Energy spectrum of gravitational waves}
 We first describe the feature in the GW spectrum expected in this scenario. The amplitude of the stochastic GWs is conventionally provided in terms of the relative spectral energy density defined as,
 	\begin{align}\label{eq:spectrum}
		\Omega_{\rm GW}(f) \equiv \frac{1}{\rho_c}\frac{{\rm d}\rho_{\rm GW}}{{\rm d}\log f}.
	\end{align}
 Assuming the primordial tensor power spectrum is almost flat in the relevant frequency band, the amplitudes for modes crossing the horizon before and after the $g_{\ast}$ change are related as \cite{Watanabe:2006qe},
 	\begin{align}
		\Omega_{\rm GW}(f_{\rm after})  =  \left(\frac{g_{\rm \ast, after}}{g_{\rm \ast, before}}\right)^{-\frac{1}{3}}\Omega_{\rm GW}(f_{\rm before}),
	\end{align}
assuming the horizon crossing occurs in the radiation-dominated era since the GW energy density $\rho_{\rm GW}$ evolves as $\propto a^{-4}$ after the horizon crossing and the critical density $\rho_c$ as $\propto g_{\ast}^{-1/3}a^{-4}$, respectively. Here, we have assumed $g_{\ast,s}=g_{\ast}$ because their difference is irrelevant to the present analysis. Then, we expect about 20\% change in the GW amplitude in this scenario. 

 The step appears at frequencies corresponding to modes that cross the horizon during $g_{\ast}$ change. For modes crossing the horizon in the radiation-dominated era, the frequency is related to the horizon-crossing temperature $T_{\rm c}$ as \cite{Watanabe:2006qe},
 	\begin{align}\label{eq:temptof}
		f = 2.7 \times 10^{-5}~{\rm Hz} \left(\frac{T_{\rm c}}{1~{\rm TeV}}\right)\left(\frac{g_{\ast}}{128}\right)^{\frac{1}{6}},
	\end{align}
where 128 corresponds to the approximate value of $g_*$ of the SM particles and the gauginos (see Fig. \ref{fig:g}).	
 Since the relevant energy range ${\cal O}(10^2-10^4)$ TeV corresponds to the frequency band of DECIGO/BBO, ${\cal O}(0.001-0.1)$ Hz, DECIGO and BBO have a potential to detect the $g_{\ast}$ change expected in this scenario.
 
 In the slow-roll inflation paradigm, the GW spectrum (\ref{eq:spectrum}) is approximated to be flat within the DECIGO/BBO band. Then, the GW spectrum can be expressed as,
 	\begin{align}\label{eq:gwspectrumform}
		\Omega_{\rm GW}(f;\theta)  =  \left[\frac{g_{\rm \ast}(T_{\rm c}; m_0)}{128}\right]^{-\frac{1}{3}}\Omega_{\rm GW,0},
	\end{align}
where $g_{\ast}(T_{\rm c}; m_0)$ is the number of relativistic degrees of freedom provided in the previous section for the scalar mass $m_0$ and $T_{\rm c}$ is related to the frequency as Eq.(\ref{eq:temptof}). 
Though the spectrum could have a tilt within the DECIGO/BBO band, we can show that our results are not affected much even if we consider the tilt of the spectrum given by the standard inflation consistency relation between the GW amplitude and the spectral index \cite{Liddle:1993fq}.

The GW spectrum (\ref{eq:gwspectrumform}) is characterized by three parameters, which we have denoted as $\theta$: the amplitude $\Omega_{\rm GW,0}$, scalar mass $m_0$, and the value of $g_{\ast}$ at high temperature $g_{\rm \ast, before}$, which respectively correspond to amplitude in the low-frequency region, frequency at the step, and amplitude in the high-frequency region. 
For convenience, we employ a normalization related to the tensor-to-scalar ratio as,
	\begin{align}\label{eq:rtoomega}
		\Omega_{\rm GW,0}  \equiv 4.4 \times 10^{-15} r_{\rm GW},
	\end{align}
though the analysis is independent of the CMB observations. 
The relation between $r_{\rm GW}$ and $r_{\rm CMB}$, the tensor-to-scalar ratio on the CMB scale, depends on the model of inflation (the primordial spectrum for tensor modes) \cite{Friedman:2006zt}. 
The value of $r_{\rm GW}$ is typically ${\cal O}(1-10)$\% less than $r_{\rm CMB}$ for relevant cases though some models predict larger values \cite{BS}. 
The relation also depends on the thermal history of the Universe. In this scenario, for example, additional entropy production is expected from decays of some meta-stable particles such as a gravitino. 
However, this contribution hardly affect the relation to an accuracy of the analysis in the typical SUSY models. 
Since $r_{\rm CMB}$ is constrained as $r_{\rm CMB} < 0.36$ at the 95\% level of confidence \cite{Komatsu:2010fb}, $r_{\rm GW}$ is expected to be ${\cal O}(0.1)$ at most. 
We also introduce the change in $g_{\ast}$ as $\Delta g_{\ast} \equiv g_{\rm \ast, before}-g_{\rm \ast, after}=g_{\rm \ast, before}-128$.
 
 In the remaining part, we investigate detectability of the step induced by the $g_{\ast}$ change with DECIGO and BBO for scalar mass $m_0$ in the range ${\cal O}(10-10^4)$ TeV.

\subsection{Detectability of the step}
 To detect a weak signal as the primordial GWs, we should perform a correlation analysis between two detectors. Here, we investigate whether DECIGO/BBO can detect the step by using the correlation analysis techniques based on Ref.  \cite{CA}.
 
\begin{figure}[htbp]
	\centering
	\includegraphics[width=.65\linewidth]{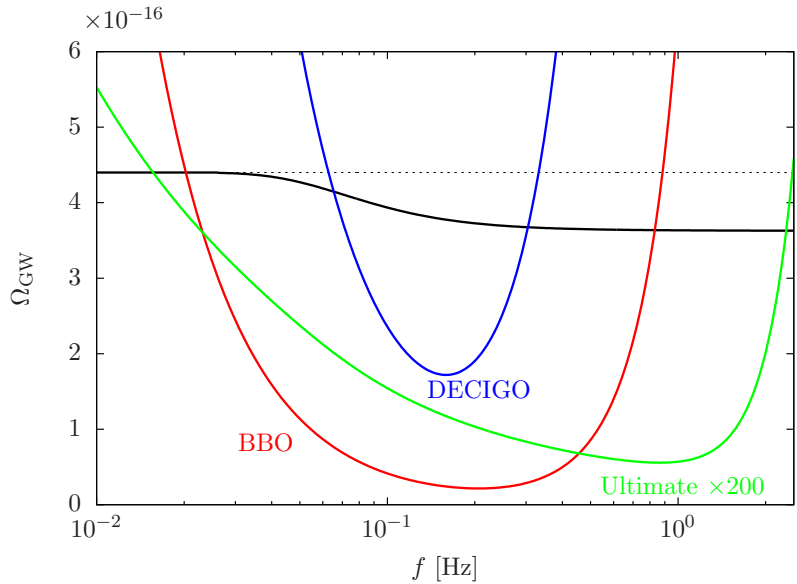}
	\caption{The GW spectrum for $m_0=10^4~{\rm TeV}$ (black line) and noise curves for DECIGO (blue), BBO (red) and ultimate-DECIGO (green).
Here we set $T_{\rm obs} = 1$ yr. 
Note that the noise (\ref{eq:gwnoise}) is determined by the contributions from GWs at lower frequencies for ultimate-DECIGO.
	}
	\label{fig:spectrum}
\end{figure}

 The output for each detector in the Fourier space $\hat{S}_i(f)~(i=1,2)$ are written in the form,
  	\begin{align}\label{eq:output}
		\hat{S}_i(f) = \hat{s}_i(f)+\hat{n}_i(f),
	\end{align}
where $\hat{n}_i(f)$ is the noise and $\hat{s}_i(f)$ is the contribution from GWs.
Dividing the frequency band into segments $F_i=[f_i-\frac{\Delta f}{2}, f_i+\frac{\Delta f}{2}]~(i=1,...,N_{\rm b})$, we consider the following combination of the outputs for two detectors,
	\begin{align}\label{eq:estimator}
		\hat{\Omega}_{\rm GW, i} \equiv \frac{1}{T_{\rm obs}\Delta f}\left(\frac{80\pi^2}{9H_0^2}\right) \int_{f \in F_i}\!{\rm d}f~f^3 \hat{S}_1^{\ast}(f)\hat{S}_2(f),
	\end{align}
where $T_{\rm obs}$ is the observation period and $H_0=73.5~{\rm km/s/Mpc}$ is the Hubble parameter at the present time \cite{Komatsu:2010fb}. Taking $\Delta f$ sufficiently small, we can evaluate the GW spectrum from the expectation value of $\hat{\Omega}_{{\rm GW}, i}$,
	\begin{align}
		\langle \hat{\Omega}_{{\rm GW}, i} \rangle \simeq \Omega_{\rm GW}(f_i),
	\end{align}
with the variance,
	\begin{align}\label{eq:gwnoise}
		\Delta {\Omega_{{\rm GW}, i}}^2 &\simeq \frac{1}{T_{\rm obs}\Delta f}\left[2\Omega_{\rm GW}(f_i)^2 + 2c_{\rm n}^2\left(\frac{80\pi^2}{9H_0^2}\right)f_i^3\Omega_{\rm GW}(f_i)S_{\rm n}(f_i) + c_{\rm n}^4\left(\frac{80\pi^2}{9H_0^2}\right)^2f_i^6S_{\rm n}(f_i)^2\right] \\ &\equiv \Delta \Omega_{\rm GW}(f_i)^2, \nonumber
	\end{align}
where $S_{\rm n}(f)$ is the square spectral noise density. Here we have introduced a factor $c_{\rm n}$ to see how improvements in the strain sensitivity, $\sqrt{S_{\rm n}}$, change the detectability. If we take the weak-signal limit, which is appropriate for near-future detectors, the expression is simplified as,
	\begin{align}\label{eq:gwnoise_weak}
		\Delta \Omega_{\rm GW}(f)^2 \simeq \frac{c_{\rm n}^4}{T_{\rm obs}\Delta f}\left(\frac{80\pi^2}{9H_0^2}\right)^2f^6S_{\rm n}(f)^2.
	\end{align}
In Fig. \ref{fig:spectrum}, we have shown the GW spectrum with $m_0=10^4~{\rm TeV}$ and $\Delta g_* = 100$ and the noise curves for DECIGO, BBO and ultimate-DECIGO \cite{Seto:2001qf, Kudoh:2005as}.
Here, we depicted $(\Delta f/f)^{1/2} \Delta \Omega_{\rm GW}$ defined in Eq. (\ref{eq:gwnoise}) as the noise and we set $T_{\rm obs} = 1$ yr. We have used fitting formulae presented in Ref. \cite{Nishizawa:2011eq} for estimating the square spectral noise density. 
In the case of ultimate-DECIGO, the noise  (\ref{eq:gwnoise}) is dominated by the contributions from GW itself at lower frequencies. 

 Detectability of the step is determined by whether the spectral shapes with different parameters, $\theta = \{ r_{\rm GW}, m_0, \Delta g_{\ast} \}$ and $\theta_f = \{ r_{{\rm GW}, f}, m_{0,f}, \Delta g_{\ast,f} \}$, can be distinguished from the data (\ref{eq:estimator}). To quantify the detectability, we introduce the following quantity,
 	\begin{align}\label{eq:measure}
		{\cal F}(\theta;\theta_f) \equiv \frac{N_{\rm d}}{\Delta f}\int_{f_{\rm cut}}\!{\rm d}f~\frac{(\Omega_{\rm GW}(f; \theta_f)-\Omega_{\rm GW}(f;\theta))^2}{\Delta \Omega_{\rm GW}(f)^2},
	\end{align}
where $N_{\rm d}$ is the number of correlation signal, which we set $N_{\rm d}=2$. 
\footnote{From Eq. (\ref{eq:measure}), we can see that $(\Delta f/f)^{1/2}\Delta \Omega_{\rm GW}$ provides effective noise in the present analysis for $\Omega_{\rm GW}$ per $\log f$, which we have plotted in Fig. \ref{fig:spectrum}}
Intuitively speaking, ${\cal F}$ represents how $\Omega_{\rm GW}(f; \theta_f)$ and $\Omega_{\rm GW}(f; \theta)$ differ from each other relative to the noise $\Delta \Omega_{\rm GW}$. 
In short, the quantity ${\cal F}$ characterizes the expectation that the true parameters $\theta_f$ are confused with $\theta$ in the future experiments. 
We use the value of ${\cal F}$ to estimate the confidence level (C.L.) in the future GW experiments.
See \ref{sec:f} for a justification.
In Eq. (\ref{eq:measure}), we have introduced a lower cut-off frequency $f_{\rm cut}$ to avoid the confusion noise from the cosmological white dwarf (WD) binaries \cite{Farmer:2003pa}. 
Strictly speaking, we should also consider contributions from the neutron star binaries. 
Here, we simply neglect them because they can be subtracted from the data for the sensitivities required for detecting the $g_{\ast}$ change as we see below \cite{Nishizawa:2011eq}. The noise introduces a cut-off in the high frequency region, $f_{\rm max}$, to Eq. (\ref{eq:measure}), though we set $f_{\rm max}=100~{\rm Hz}$ in the estimation.
Note that ${\cal F}$ does not depend on the bin size $\Delta f$. Thus, the value of $\Delta f$ is irrelevant to the analysis if it is sufficiently small. We discuss the detectability by evaluating ${\cal F}$ for $\Delta g_{\ast,f} = 100$ and $\Delta g_{\ast}=0$.

\begin{figure}[htbp]
	\centering
	
	\subfigure[DECIGO]{
	\includegraphics[width=.48 \linewidth]{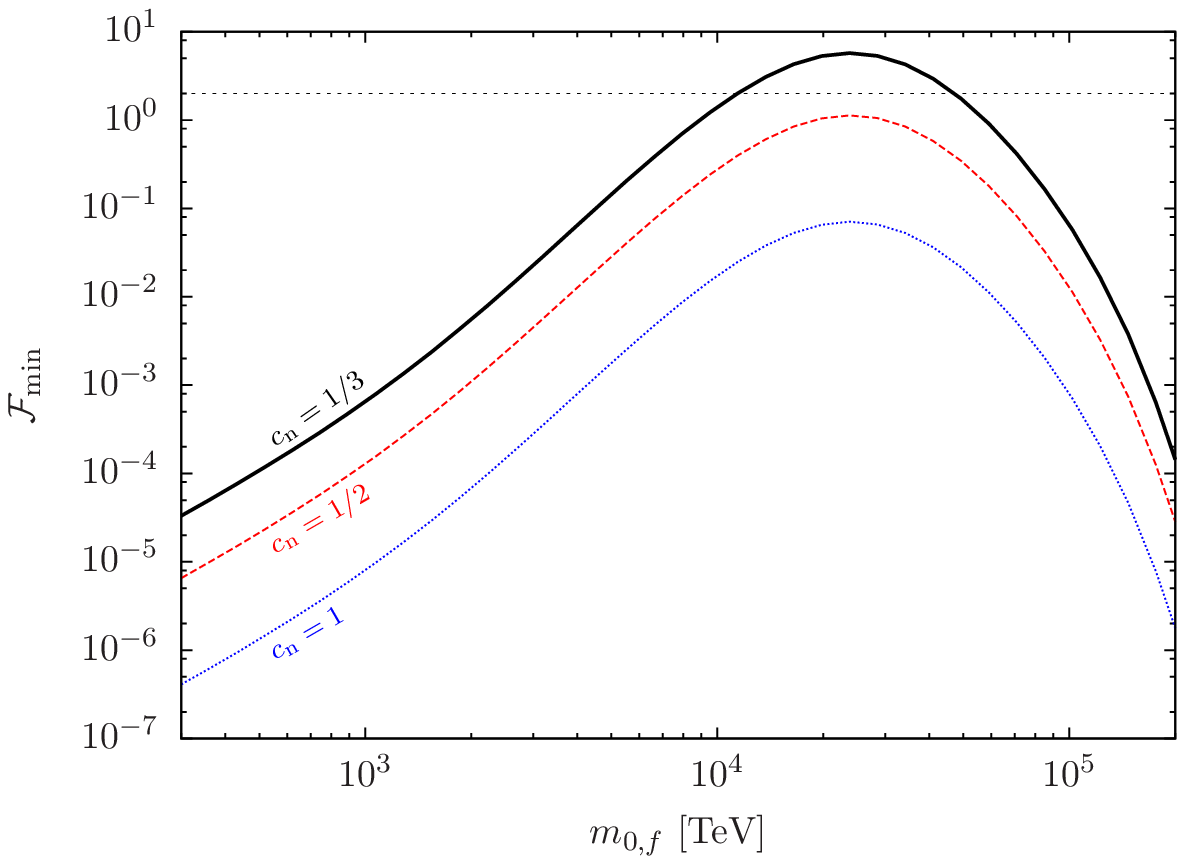}}
	\subfigure[BBO]{
	\includegraphics[width=.48\linewidth]{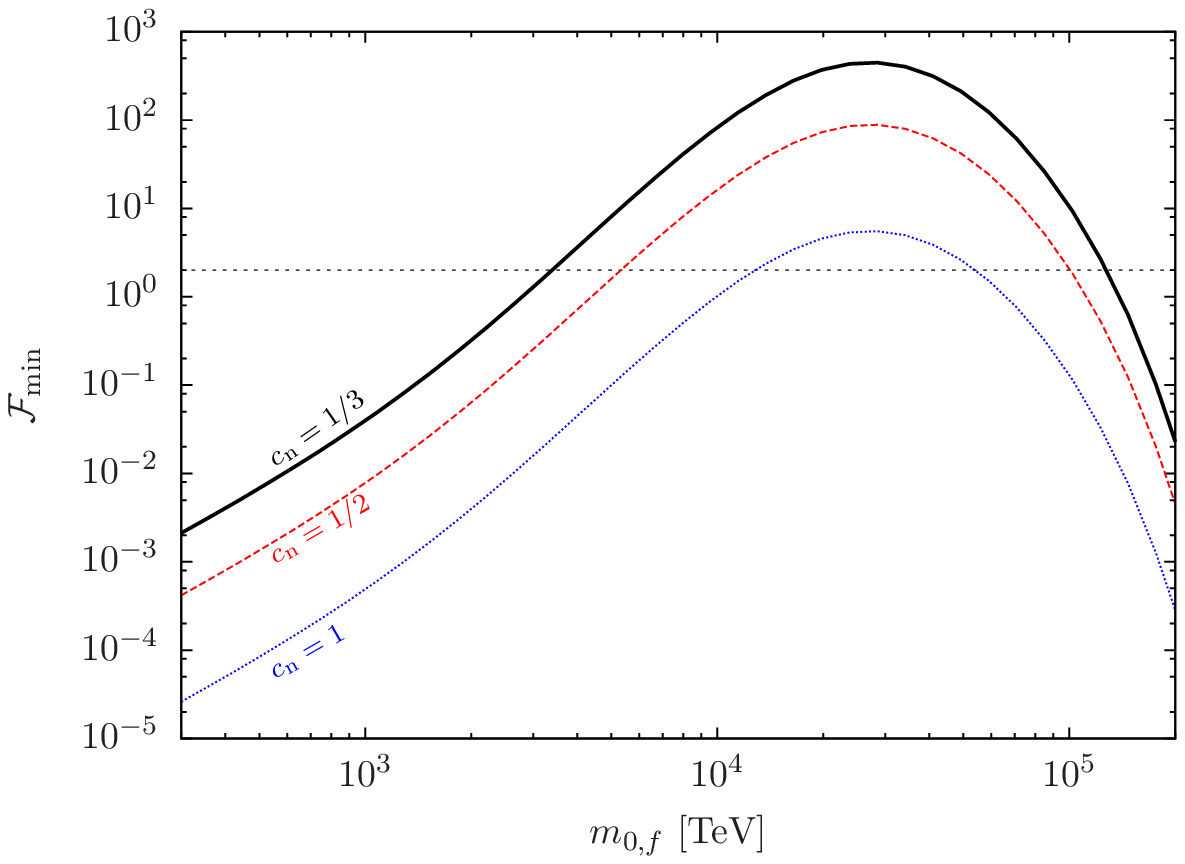}
	}
	\caption{The measure ${\cal F}$ minimized with respect to $r_{\rm GW}$ for DECIGO (left) and BBO (right) for the improvement factor $c_{\rm n}=1,1/2,1/3$ from bottom to top. 
	Here, we have set $r_{{\rm GW},f}=0.1$, $T_{\rm obs}=5~{\rm yr}$, and $f_{\rm cut}=0.1~{\rm Hz}$ for both detectors. The horizontal line indicates ${\cal F}_{\rm min}=2$.}
	\label{fig:mincn}
\end{figure}

\begin{figure}[htbp]
	\centering
	\subfigure[DECIGO with $c_{\rm n} = 1/3$]{
	\includegraphics[width=.46 \linewidth]{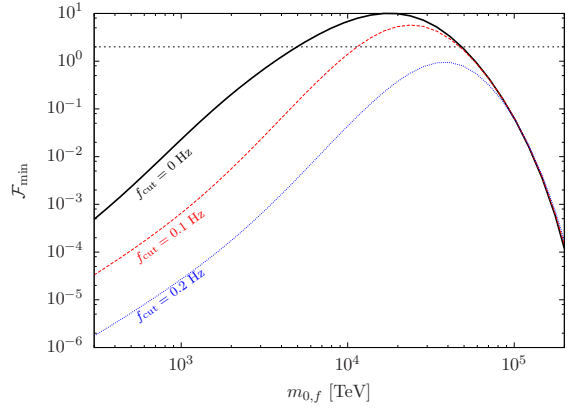}
	}
	
	\subfigure[BBO with $c_{\rm n} = 1/3$]{
	\includegraphics[width=.46\linewidth]{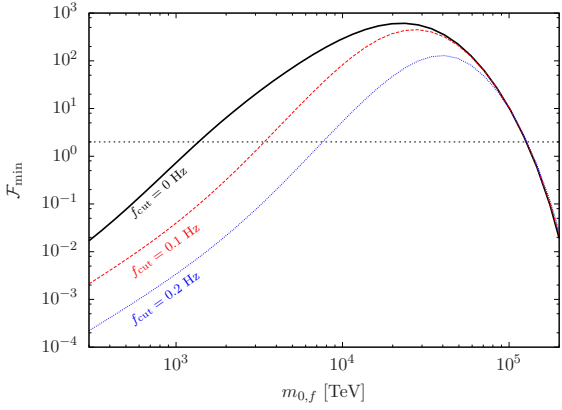}
	}
	\subfigure[Ultimate-DECIGO]{
	\includegraphics[width=.46\linewidth]{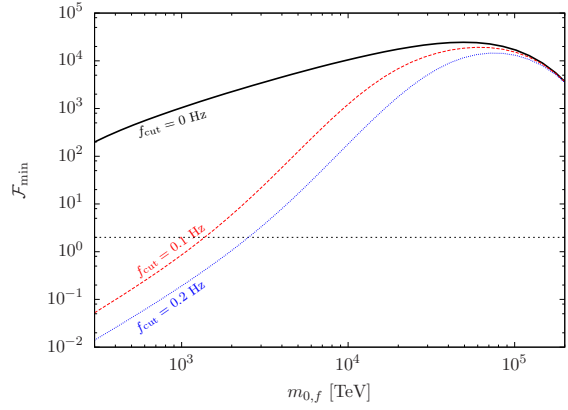}
	}
	\caption{The measure ${\cal F}$ minimized with respect to $r_{\rm GW}$  for the lower cut-off frequency $f_{\rm cut}=0~{\rm Hz}, 0.1~{\rm Hz}, 0.2~{\rm Hz}$. 
	Here, we have set $r_{{\rm GW},f}=0.1$, $c_{\rm n}=1/3$, and $T_{\rm obs}=5~{\rm yr}$. The horizontal line indicates ${\cal F}_{\rm min}=2$. We have also depicted ${\cal F}_{\rm min}$ for ultimate-DECIGO.}
	\label{fig:minfcut}
\end{figure}

 In Fig. \ref{fig:mincn}, we have depicted ${\cal F}$ minimized with respect to $r_{\rm GW}$, \footnote{Note that ${\cal F}_{\rm min}$ is independent of $m$ because the GW spectrum is flat for $\Delta g_{\ast}=0$. Therefore, ${\cal F}_{\rm min}$ is given by a function of $m_f$ for $\Delta g_{\ast,f} = 100$ and $\Delta g_{\ast}=0$.} ${\cal F}_{\rm min}$, for  $\Delta g_{\ast,f} = 100$ and $\Delta g_{\ast}=0$. 
 Here, ${\cal F}$ is minimized to take into account that degeneracy with $r_{\rm GW}$ increases estimation error.
 The curves are presented for various values of the improvement factor $c_{\rm n}=1,1/2,1/3$. 
 In the estimation, we have set the other parameters as $r_{{\rm GW},f}=0.1$, $T_{\rm obs}=5~{\rm yr}$, and $f_{\rm cut}=0.1~{\rm Hz}$. 
 Note that the values of ${\cal F}_{\rm min}$ for another parameter choice can be easily obtained by scaling ${\cal F}_{\rm min}$ as $ \propto T_{\rm obs}r_{{\rm  GW}, f}^2c_{\rm n}^{-4}$ in the weak-signal limit. 
 We have also shown the curves for different lower cut-off frequencies $f_{\rm cut}$ in Fig. \ref{fig:minfcut} to see the effect of the WD noise for the estimation. 
 We assume $r_{{\rm GW},f}=0.1$ and $T_{\rm obs}=5~{\rm yr}$ and set $c_{\rm n}=1/3$ for DECIGO and BBO.
 To see a detectability in an ideal case, we have also depicted ${\cal F}_{\rm min}$ for ultimate-DECIGO. In this case, the variance (\ref{eq:gwnoise}) is dominated by the contributions from GWs and the detector noise $S_{\rm n}$ only determines an effective cut-off in the high frequency region. Hence, the values of ${\cal F}_{\rm min}$ are approximately constant for different values of $r_{{\rm GW}, f}$ as far as the contributions from the detector noise can be neglected.
 
  From the figures, we can see that BBO can test our scenario with $m_0 = {\cal O}(10^3-10^4)~{\rm TeV}$ if $r_{\rm GW}$ is large enough, though testability in the low mass region depends on the lower cut-off frequency associated with the WD noise. 
  Though sensitivity of the standard DECIGO is not enough to detect the $g_{\ast}$ change \cite{Chiba:2007kz}, DECIGO can be comparable with BBO if we employ an improved design with three times better sensitivity.  
  A degree of improvement in sensitivity is important to extend the testable mass range. 
  A large region in the mass range ${\cal O}(10^3-10^4)~{\rm TeV}$ can be covered by the interferometers as BBO-grand ($c_{\rm n} \simeq 1/3$) or ultimate-DECIGO, though the WD noise should be removed to test the lower mass region $m_0 < {\cal O}(10^3)$ TeV.  

\begin{figure}[htbp]
	\centering
	\begin{minipage}{.48\linewidth}
	\includegraphics[width=\linewidth]{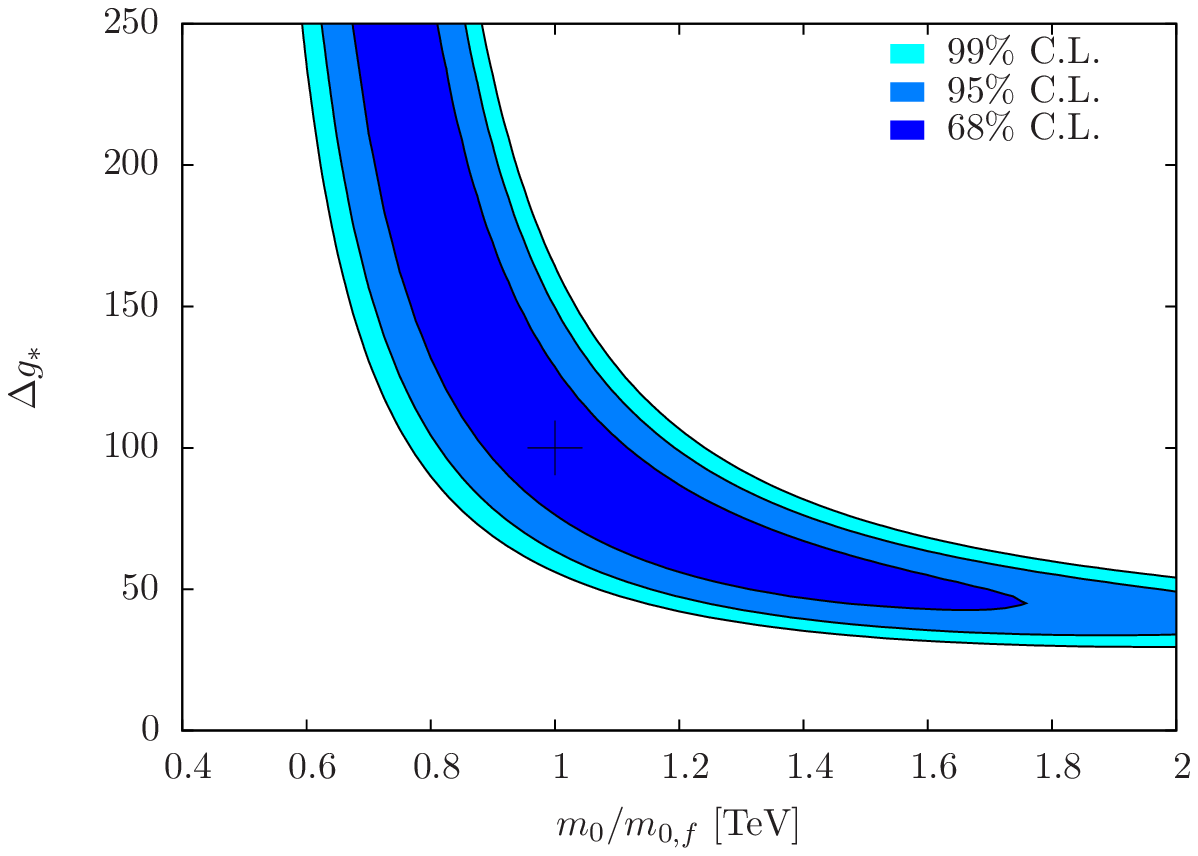}
	\end{minipage}
	\begin{minipage}{.48\linewidth}
	\includegraphics[width=\linewidth]{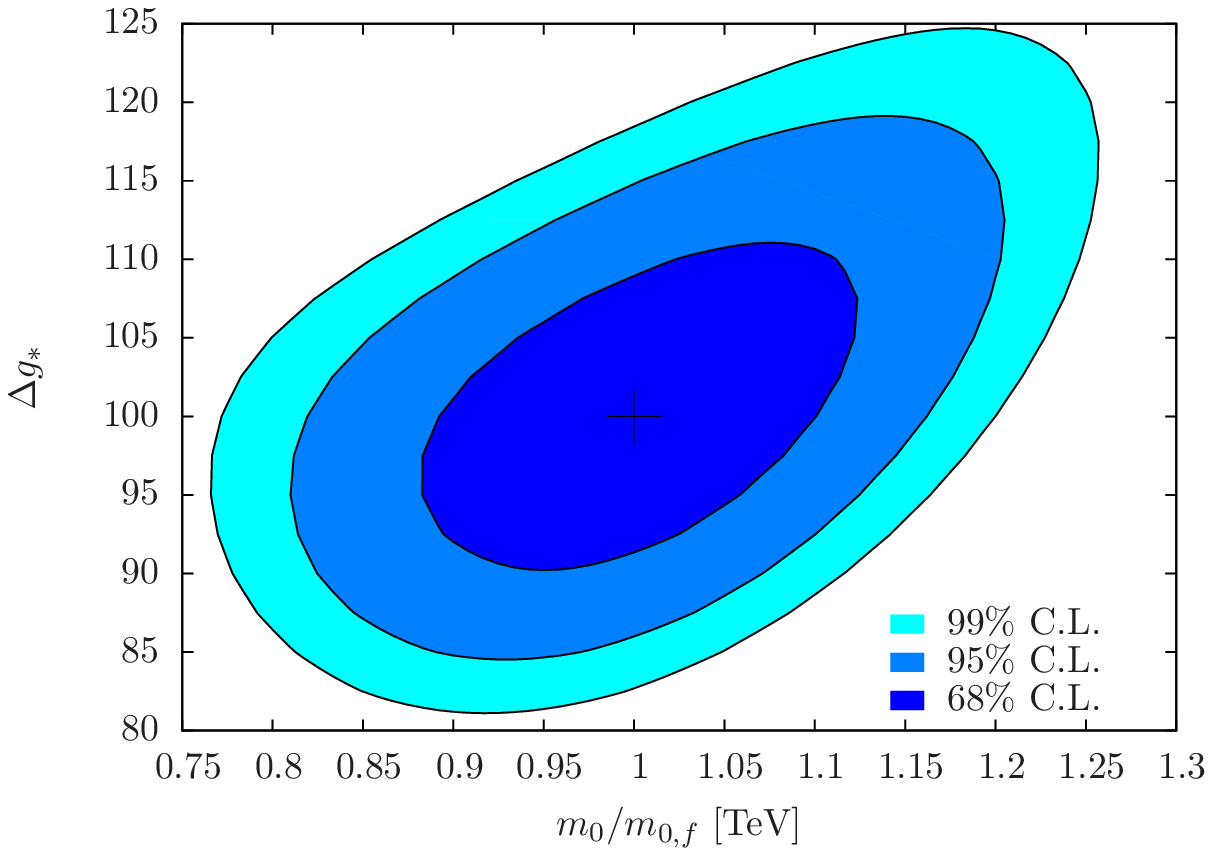}
	\end{minipage}	
	\caption{Expected error in $(m_0, \Delta g_{\ast})$ with $(r_{{\rm GW}, f}, m_{0,f}, \Delta g_{\ast,f})=(0.1, 10^4~{\rm TeV}/3\times 10^4~{\rm TeV}, 100)$ for BBO with $c_{\rm n}=1/3$ (left: $10^4$~{\rm TeV}, right: $3\times 10^{4}~{\rm TeV}$). Here, ${\cal F}$ is minimized with respect to $r_{\rm GW}$ to take into account that degeneracy with $r_{\rm GW}$ increases estimation error for $(m_0, \Delta g_{\ast})$. Here, we have set $T_{\rm obs}=5~{\rm yr}$ and $f_{\rm cut}=0.1~{\rm Hz}$.}
	\label{fig:contours}
\end{figure}

In Fig. \ref{fig:contours}, we have also depicted expected error \footnote{See \ref{sec:f} for the precise meaning of the confidence level depicted here.} in $(m_0, \Delta g_{\ast})$ for BBO with $c_{\rm n}=1/3$ from the value of ${\cal F}_{\rm min}$, assuming $(r_{{\rm GW}, f}, m_{0,f}, \Delta g_{\ast,f})=(0.1, 10^4~{\rm TeV}/3\times10^4~{\rm TeV}, 100)$. 
The other parameters have been set as $T_{\rm obs}=5~{\rm yr}$ and $f_{\rm cut}=0.1~{\rm Hz}$. 
Note again that the value of ${\cal F}_{\rm min}$ for another parameter choice is obtained by scaling ${\cal F}_{\rm min} $ as $\propto T_{\rm obs}r_{{\rm  GW}, f}^2c_{\rm n}^{-4}$ in the weak signal limit. 
From the figures, we can see the interferometers as BBO-grand have a potential to determine the scalar mass scale, though precise value of error depends on the MSSM spectrum. 
The parameter degeneracies are observed for $m_{0,f}=10^4~{\rm TeV}$ because the WD noise covers the spectrum in the lower frequency region.
 
\section{Conclusion and Discussion}
In this paper, we have discussed a GW test of the SSMs with heavy scalar particles motivated with the recent LHC Higgs search.
We have quantitatively shown that the future GW experiments as DECIGO and BBO have a potential to test the heavy scalar scenario, $m_0 ={\cal O}(10^3 - 10^4) $ TeV, and determine the scalar mass scale. 
Even in the case that the GW amplitude is large, a degree of improvement in sensitivity is necessary to cover a large region in the mass range ${\cal O}(10^3-10^4)~{\rm TeV}$ by DECIGO and BBO. We have seen that it can be covered by using BBO-grand or ultimate-DECIGO.
Our analysis depends on the assumption for the spectral shape of the primordial GW spectrum in the DECIGO/BBO band as is naturally expected. For example, if the reheating temperature to evade the gravitino problem is too low, the spectrum is expected to be deformed in the DECIGO/BBO band \cite{RH}. 
However, in the case of very high reheating temperature scenario, such as the thermal leptogenesis scenario,
we can safely make the analysis assuming the flat spectrum. 
In addition, the spectrum could have a tilt in the DECIGO/BBO band. Although it can be shown that our results do not change much under the standard inflation consistency relation, introduction of the tilt could increase error in the estimation much if no prior knowledge on the tilt is assumed. In that case, better sensitivity in the high frequency region will be needed to break the degeneracy of the models.

 \section*{Acknowledgement}
 We would like to thank A. Nishizawa for reading the manuscript and T. Tanaka for useful comments. We would also like to thank T. T. Yanagida for his advice and encouragement and K. Nakayama for pointing out our numerical error in Eq. (\ref{eq:temptof}). The work of RS is supported by a Grant-in-Aid through JSPS.

\appendix
\gdef\thesection{Appendix \Alph{section}}
\section{The meaning of ${\cal F}$} \label{sec:f}
 Here, we clarify the meaning of ${\cal F}$ given by Eq. (\ref{eq:measure}). Provided data $\hat{\Omega}_{{\rm GW},i}$, the parameters $\theta$ can be estimated by maximizing a likelihood function provided as,
 	\begin{align}\label{eq:likelihood}
		{\cal L}(\theta; \hat{\Omega}_{\rm GW}) &\equiv \frac{1}{{\cal N}}\exp\left[-\frac{N_{\rm d}}{2}\sum_i~\frac{(\hat{\Omega}_{{\rm GW},i}-\Omega_{\rm GW}(f_i;\theta))^2}{\Delta \Omega_{{\rm GW},i}^2}\right] \nonumber \\
		&\simeq \frac{1}{{\cal N}}\exp\left[-\frac{1}{2}\cdot\frac{2N_{\rm d}}{\Delta f}\int_{f_{\rm cut}}\!{\rm d}f~\frac{(\hat{\Omega}_{\rm GW}(f)-\Omega_{\rm GW}(f;\theta))^2}{\Delta \Omega_{\rm GW}(f)^2}\right],
	\end{align}
where ${\cal N}$ is a normalization constant.
\footnote{A factor 2 appears in Eq. (\ref{eq:likelihood}) because Fourier modes are defined in $(-\infty, \infty)$.} 
Here, we have assumed that each segment $F_i$ contains so large number of Fourier modes that $\hat{\Omega}_{{\rm GW},i}$ can be considered to be Gaussian distributed. This assumption can be compatible with small bin size because the frequency resolution, $T_{\rm obs}^{-1} \sim {\cal O}(10^{-8})~{\rm Hz}$, is much smaller than the relevant frequency, ${\cal O}(0.1)~{\rm Hz}$. Then, the estimated value of parameters $\theta$, maximum likelihood estimators (MLE), are obtained by maximizing the likelihood function (\ref{eq:likelihood}), or minimizing the quantity,
\begin{align}
	\chi^2 \equiv \frac{2N_{\rm d}}{\Delta f}\int_{f_{\rm cut}}\!{\rm d}f~\frac{(\hat{\Omega}_{\rm GW}(f)-\Omega_{\rm GW}(f;\theta))^2}{\Delta \Omega_{\rm GW}(f)^2},
\end{align}
and the error (confidence level) in the parameters by seeing the value of $\Delta \chi^2 \equiv \chi^2-\chi_{\rm min}^2$.

 Now we consider the expectation that parameters $\theta$ are judged to be consistent by future experiments when true values of parameters are $\theta_f$: $\langle \hat{\Omega}_{\rm GW}(f) \rangle = \Omega_{\rm GW}(f;\theta_f)$. 
 To estimate the accuracy of the estimation, we should consider not only the error expected in the experiments but also deviation of MLE from the true values, which depends on the data obtained in the future. 
 This uncertainty of MLE can be included by taking the ensemble average
 over realizations of the data $\hat{\Omega}_{\rm GW}(f)$ with $\langle \hat{\Omega}_{\rm GW}(f) \rangle = \Omega_{\rm GW}(f;\theta_f)$:
 	\begin{align}\label{eq:exlikelihood}
		\langle {\cal L}(\theta; \hat{\Omega}_{\rm GW}) \rangle &\propto \exp\left[-\frac{1}{2} \cdot \frac{N_{\rm d}}{\Delta f}\int_{f_{\rm cut}}\!{\rm d}f~\frac{(\Omega_{\rm GW}(f;\theta_f)-\Omega_{\rm GW}(f;\theta))^2}{\Delta \Omega_{\rm GW}(f)^2}\right].
	\end{align}
 The exponent is just $-{\cal F}/2$. The quantity ${\cal F}$ measures the compatibility of the future data with the model where values of parameters are $\theta$ when the true values are $\theta_f$. 
 
\begin{figure}[htbp]
	\centering
	\includegraphics[width=1\linewidth,angle=0]{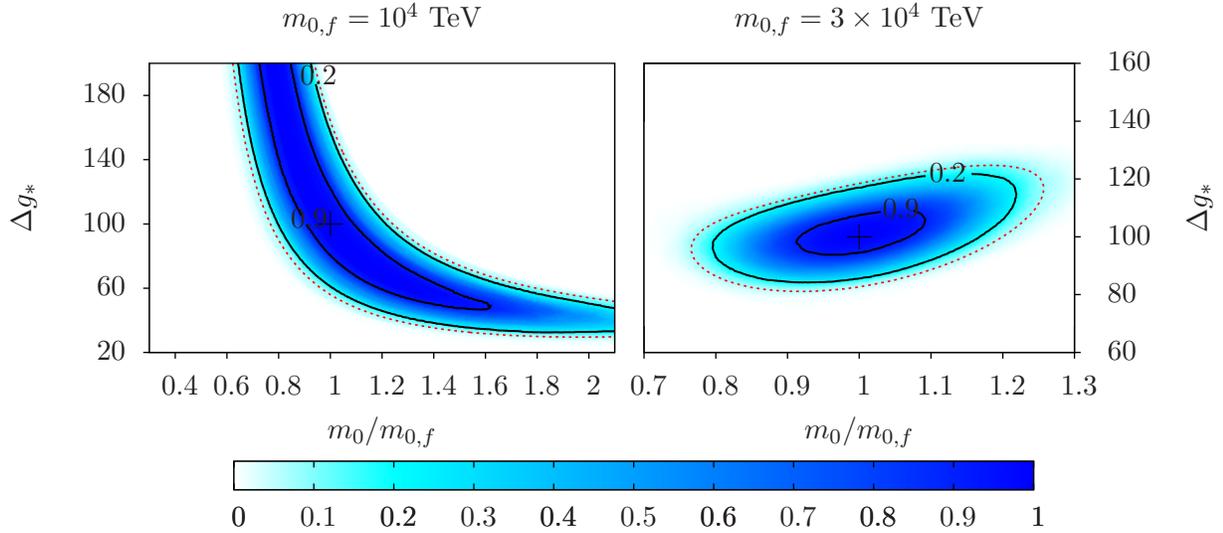}

	\caption{The result of Monte Carlo simulation for parameter regions where parameters are expected to be in the 99\% C.L. contour in the future GW experiment.
			The color panel shows the probability that parameters are contained in 99\% C.L. contour. The dashed and red line represents ``99\% C.L. contour" depicted by using the value of ${\cal F}$ and the solid and black lines the contours with a constant probability. To depict the figure, we have used BBO with $c_{\rm n}=1/3$, $T_{\rm obs}=5~{\rm yr}$, $f_{\rm cut}=0.1~{\rm Hz}$, and $(r_{{\rm GW}, f}, m_{0,f}, \Delta g_{\ast,f})=(0.1, 10^4~ {\rm TeV}/ 3\times 10^4~{\rm TeV}, 100)$. (left: $10^4$~{\rm TeV}, right: $3\times 10^{4}~{\rm TeV}$)}
	\label{fig:MC}
\end{figure}

 We can relate ${\cal F}$ to the confidence level expected in the future GW experiments as $e^{-{\cal F}/2} = 2^{N_{\rm d}/2}\langle e^{-\chi_{\rm min}^2/2}e^{-\Delta \chi^2/2} \rangle$. Here, the average of $\Delta \chi^2$ is taken with a weight $e^{-\chi_{\rm min}^2/2}$, the maximum value of the likelihood function, which measures the goodness-of-fit of our model with a realization of the data. 
 Hence, ${\cal F}$ approximates $\Delta \chi^2$ averaged over realizations except those less compatible with our model, for which the estimation of parameters is less reliable. 
  In Fig. \ref{fig:MC}, we have depicted the result of Monte Carlo simulation for parameter regions where parameters are expected to be in the 99\% C.L. contour with a given probability, assuming $(r_{{\rm GW}, f}, m_{0,f}, \Delta g_{\ast,f})=(0.1, 10^4~{\rm TeV}/3\times10^4~{\rm TeV}, 100)$ for BBO with $c_{\rm n}=1/3$, $T_{\rm obs}=5~{\rm yr}$, and $f_{\rm cut}=0.1~{\rm Hz}$. From the figure, we can see that the ``99\% C.L. contour" depicted by using the value of ${\cal F}$ covers almost all of parameters that can have the 99\% C.L. in the future analysis; the probability that parameters outside the contour have 99\% C.L. is less than $\sim$ 5\%. Therefore, we use the value of ${\cal F}$, which can be estimated analytically, as a measure for the expected error of the parameter estimation in the future GW experiments.
  
\end{document}